\documentclass[prl,reprint,amsmath,amssymb,floatfix,citeautoscript]{revtex4-1}
\usepackage{cancel}
\usepackage{amsmath}
\usepackage{physics}
\usepackage{graphicx}
\usepackage{amssymb}
\usepackage{amsthm}
\usepackage{bm}
\usepackage{dcolumn}
\usepackage{braket}
\usepackage{longtable}

\usepackage{ragged2e}
\usepackage{txfonts}

\usepackage{color}
\usepackage[usenames,dvipsnames]{xcolor}
\definecolor{myblue}{rgb}{0,0,1}
\usepackage[breaklinks=true,colorlinks=true,linkcolor=myblue,urlcolor=myblue,citecolor=myblue]{hyperref}

\let\vr\undefined
\newcommand{\vr}{{\bm{r}}}

\newcommand{\vq}{{\bm{q}}}
\newcommand{\vk}{{\bm{k}}}

\newcommand{\eps}{{\varepsilon}}

\begin{document}

\title{
Ab Initio Lifetime and Concomitant Double-Excitation Character
of Plasmons at Metallic Densities}

\author{Alan M. Lewis}
\affiliation{Department of Chemistry and James Franck Institute,
University of Chicago, Chicago, Illinois 60637, USA}
\author{Timothy C. Berkelbach}
\affiliation{Department of Chemistry, Columbia University, New York, New York 10027 USA}
\affiliation{Center for Computational Quantum Physics, Flatiron Institute, New York, New York 10010 USA}
\email{tim.berkelbach@gmail.com}

\begin{abstract}
The accurate calculation of excited state properties of interacting electrons
in the condensed phase is an immense challenge in computational physics.
Here, we use state-of-the-art equation-of-motion coupled-cluster theory with single
and double excitations (EOM-CCSD) to calculate the dynamic structure
factor, which can be experimentally measured by inelastic x-ray and electron scattering.  
Our calculations are performed on the uniform electron gas at densities
corresponding to Wigner-Seitz radii of $r_s=5$, 4, and 3 corresponding to the
valence electron densities of common metals.  We compare our
results to those obtained using the random-phase approximation, which is known
to provide a reasonable description of the collective plasmon excitation and
which resums only a small subset of the polarizability diagrams included in
EOM-CCSD.  We find that EOM-CCSD, instead of providing a perturbative
improvement on the RPA plasmon, predicts a many-state plasmon resonance, where
each contributing state has a double-excitation character of 80\% or more. 
This finding amounts to an ab initio treatment of the plasmon linewidth, which
is in good quantitative agreement with previous diagrammatic calculations,
and highlights the strongly correlated nature of lifetime effects in
condensed-phase electronic structure theory.
\end{abstract}

\maketitle

\textit{Introduction.} The uniform electron gas (UEG) is a paradigmatic model of
interacting electrons in the condensed
phase~\cite{PinesNozieres,GiulianiVignale}.  Finite-order perturbation theory
for the ground-state correlation energy exhibits a divergence due to the UEG's
metallic character and long-ranged Coulomb interactions.  These divergences are
famously removed by the infinite-order resummation of time-independent
particle-hole ring diagrams known as the random-phase approximation
(RPA)~\cite{Bohm1953,GellMann1957}.  As a dynamical theory of the density
response function, the RPA corresponds to a resummation of all time-dependent
ring diagrams and forms the microscopic basis for screening the Coulomb
interaction, as is done, for example, in the $GW$
approximation~\cite{Hedin1965}.  In the UEG, the RPA strongly modifies the
noninteracting polarizability, most significantly predicting the existence of a
collective plasmon excitation. Outside of the particle-hole continuum, the RPA
plasmon is a coherent, dispersive excitation with infinite lifetime; inside the
particle-hole continuum, it acquires a lifetime due to Landau damping.

\begin{figure}[b]
\centering
\includegraphics[scale=1.0]{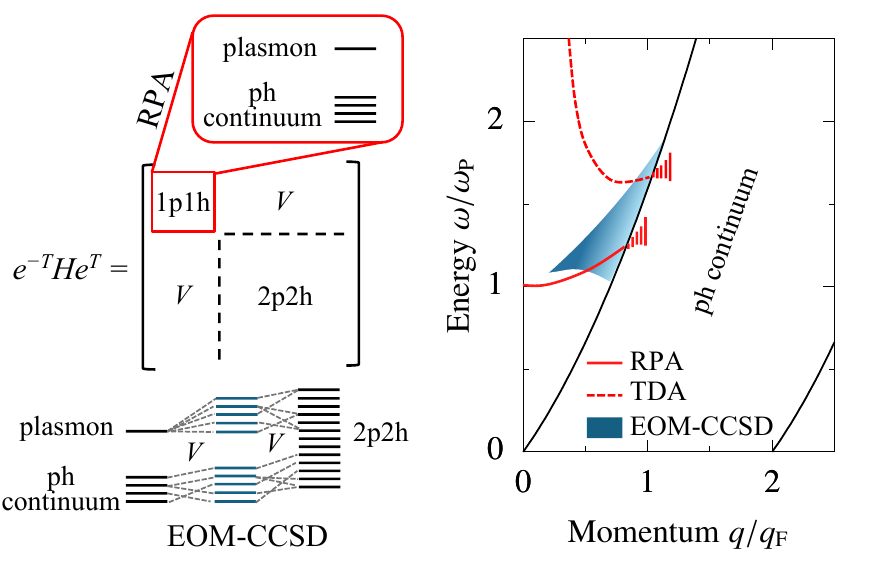}
\caption{
Schematic of the main methods and results of the this work.  
EOM-CCSD diagonalizes a similarity-transformed Hamiltonian in the
1p1h and 2p2h configuration space; diagonalization within the 1p1h
space gives the RPA result (plus minor corrections~\cite{Berkelbach2018}),
which yields a collective plasmon split off from the remaining ph continuum. 
The interaction of these states with 2p2h configurations produces
new eigenstates, as shown in the bottom left. 
The single plasmon state is mixed into many eigenstates, giving it an effective
lifetime, although each individual eigenstate has predominantly 2p2h character.
As shown at the right, this leads to a plasmon dispersion that is slightly
higher than that of the RPA, in contrast to the known exact behavior,
but with a proper correlation-induced lifetime.
}
\label{fig:scheme}
\end{figure}

A number of calculations have attempted to improve on the RPA treatment of the
UEG density response function, in order to uncover signatures of electron
correlation and to test new theoretical tools.
Previous works extend the RPA through the selective inclusion of a
static or dynamic local field 
correction~\cite{Singwi1968,Singwi1970,Vashishta1972,Kugler1975,
Mukhopadhyay1975,Dabrowski1986,Qian2002}, 
which is closely related to time-dependent
density functional theory.
However, fully ab initio nonperturbative techniques are now reaching a point of maturity
where they can be applied in an unbiased manner to large, condensed-phase systems.
Here, we use
equation-of-motion coupled-cluster theory with single and double excitations
(EOM-CCSD)~\cite{Koch1990,Stanton1993,Kobayashi1994,Bartlett2007,Krylov2008} 
to calculate the density response function of the UEG and to
compare to that predicted by the RPA.  As shown
previously~\cite{Emrich1981,Berkelbach2018}, EOM-CCSD rigorously resums a larger
class of time-dependent diagrams than those included in the RPA.  In particular,
beyond the RPA ring diagrams, the EOM-CCSD response function includes all ladder
diagrams, mixed ring-ladder diagrams, and exchange diagrams, as well as large
classes of diagrams associated with two particle-hole pairs
(i.e.~double excitations in the excited-state wavefunction).  The
CC formalism is appropriate for periodic systems because it has
total energies that are size-extensive~\cite{Bartlett2007}.
Recent years have seen intense activity on the applications of perturbation
theory, RPA, and CC theory to realistic materials, with encouraging
results~\cite{Schimka2010,Olsen2011,Paier2012,Booth2013,Shepherd2013,Gruneis2014,Gruneis2015,McClain2016,Gruber2018}.
However, excited-state CC techniques~\cite{Koch1990,Krylov2008} have been studied
significantly less in the solid-state, despite their attractive properties.
For example, unlike quantum Monte Carlo approaches, EOM-CCSD directly constructs
spectral functions on the real frequency axis, and does not require analytic
continuation.  To the best of our knowledge, this work represents the first
calculation of neutral excitation spectra of a periodic, condensed-phase 
system using EOM-CCSD.  
This technological advance opens the door for highly accurate simulations of
excitonic and plasmonic phenomena in atomistic materials. 

\textit{Theory.}
Our calculations are performed in a finite simulation cell with periodic
boundary conditions in the canonical ensemble at zero temperature.  
The simulation cell contains $N$ electrons in a volume
$\Omega = L^3$. The electron density is $n=N/\Omega$ and the Wigner-Seitz
radius, measuring the average size of a sphere occupied by one electron, is $r_s
= [3/(4\pi n)]^{1/3}$.  The pairwise Coulomb interaction is treated with the
periodic Ewald interaction~\cite{Fraser1996,Chiesa2006,Drummond2008} assuming a
compensating background charge.
The one-electron basis functions are plane waves, $\psi_\vk(\vr) = \Omega^{-1/2}
\exp(i\vk\cdot\vr)$ where $\vk = (2\pi/L) [l,m,n]$ and $l,m,n$ are integers.
For closed-shell electron configurations, the spectrum of the finite UEG is
always gapped, but becomes increasingly metallic at large system sizes.  
In the correlated calculations to follow, we use $N=66$ electrons
in a single-particle basis of 81 plane-wave orbitals, corresponding to a
density-dependent kinetic energy cutoff.  This system size is comparable to
those used in ground-state quantum Monte Carlo~\cite{Ceperley1980,Moroni1995,Kwon1998,Kwee2008,Drummond2009,Dornheim2016}
and quantum
chemistry~\cite{Shepherd2012FCIQMC,Shepherd2012CBS,Gruneis2013,Shepherd2014,Shepherd2014PRL,Shepherd2016,Spencer2016}
calculations on the UEG, for which finite-size and basis set effects have been studied more methodically.  
The computational cost of these CCSD calculations scales like $N^2 M^4$.
For these parameters, the
ground-state CCSD calculation, which is only performed once for each value of
$r_s$, is relatively cheap.  Although EOM-CCSD has the same formal scaling, the
calculation of the dynamic structure factor is our bottleneck, because of the
many frequency points and challenges associated with the solution of a system of
linear equations.  We have performed simple finite-size analysis and believe
that our conclusions remain valid in the thermodynamic limit.

Our primary observable is the dynamic structure factor 
$S(\vq,\omega)=-\pi^{-1} \mathrm{Im} \Pi(\vq,\omega)$,
where $\Pi(\vq,\omega)$ is the polarizability, i.e.~the Fourier
transform of the retarded density response function,
\begin{align}
\Pi(\vq,\omega) &= -i\int_0^\infty dt e^{i\omega t} 
    \langle \Psi_0 | [ 
        \rho(\vq,t), \rho^\dagger(\vq,0) ] | \Psi_0 \rangle,
\end{align}
with $\rho^\dagger(\vq) = \sum_\vk a^\dagger_{\vk+\vq} a_{\vk}$.
For the noninteracting electron gas, the polarizability can be simply
computed, $\Pi^0(\vq,\omega) = \sum_\vk [n_\vk - n_{\vk+\vq}]/
    [\hbar\omega - (\eps_{\vk+\vq}-\eps_\vk) + i\eta]$.
The noninteracting structure factor has a particle-hole continuum
with boundaries determined by the Fermi occupancy functions $n_\vk$. 
The exact polarizability can be formally given by 
\begin{equation}
\Pi(\vq,\omega) = \frac{\Pi^{0}(\vq,\omega)}
    {1 - v(q)[1-G(\vq,\omega)]\Pi^{0}(\vq,\omega)},
\end{equation}
where $G(\vq,\omega)$ is a dynamic local field factor.  The RPA
polarizability, which resums all time-dependent ring diagrams, is
obtained for $G(\vq,\omega)=0$ and
exhibits a pole associated with the collective, plasmon excitation.
As long as the plasmon energy falls outside of the particle-hole continuum, the
RPA predicts it to have an infinite lifetime; in other words, even in the
thermodynamic limit, the RPA plasmon is a \textit{single quantum state}.  Inside
the particle-hole continuum, the plasmon interacts with quasiparticle
excitations, leading to Landau damping and a finite lifetime.  In the
long-wavelength limit, the RPA is exact~\cite{GiulianiVignale} and the plasmon
dispersion approaches the classical plasma energy
$\omega_\mathrm{P}(q\rightarrow 0) = \sqrt{4\pi n}$.  At finite $q$, the exact
plasmon dispersion is unknown; however, by analyzing the limiting behaviors and
conservation laws, one can argue that the exact plasmon dispersion lies below
that predicted by the RPA~\cite{GiulianiVignale}.  Therefore, an exact treatment
of the UEG is expected to produce a plasmon with a displacement to lower
energies and with a finite lifetime, both due to interaction with multipair
excitations beyond the RPA.

\begin{figure*}[t]
\centering
\includegraphics[scale=1.0]{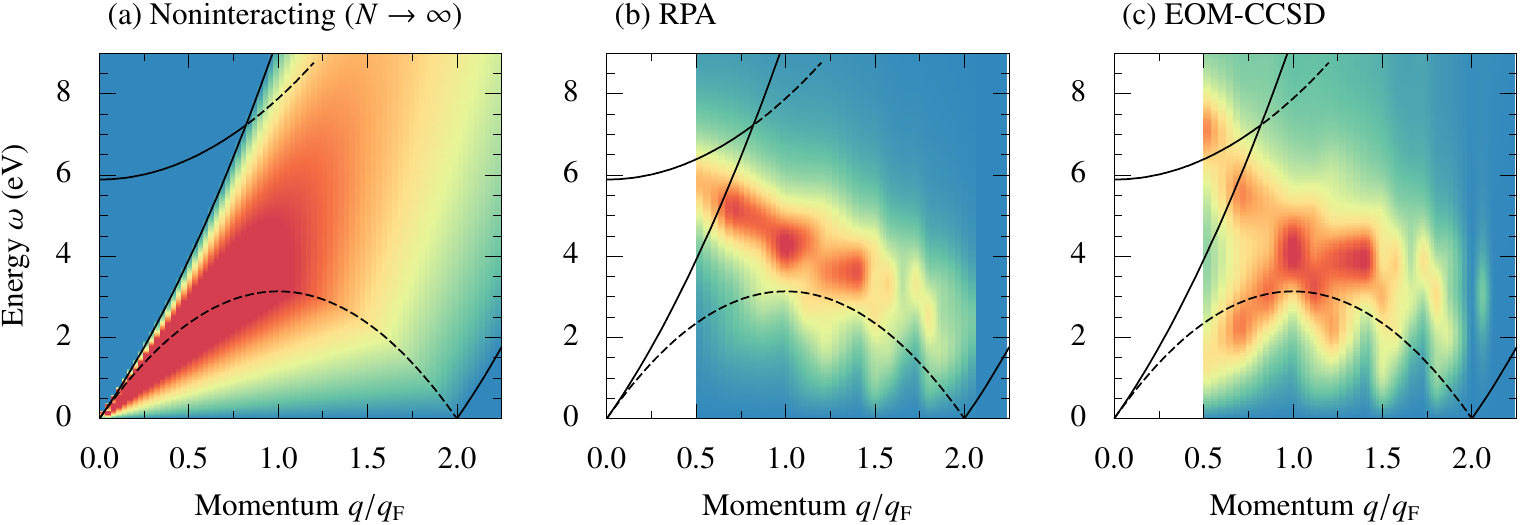}
\caption{
Dynamic structure factor of the uniform electron gas at $r_s = 4$ obtained in
the thermodynamic limit for the noninteracting theory (a) and for the finite
system from the RPA (b) and EOM-CCSD theory (c). Finite-sized calculations have
$N=66$~electrons with $M=81$ plane-wave basis functions.  Solid and dashed black
lines indicate the boundaries and maximum intensity of independent-particle
excitations and the dispersion of the plasmon calculated in the RPA. A
broadening of $\eta = 1$~eV is used in the RPA and EOM-CCSD calculations.
}
\label{fig:sqw_2d_rs4}
\end{figure*}

In recent work, one of us (T.C.B.) showed that the polarizability diagrams
summed in the RPA are a strict subset of those included in the EOM-CCSD
polarizability~\cite{Berkelbach2018}.  
However, because our analysis will make use of the
EOM-CCSD excited-state wavefunctions, we briefly review the plasmon
wavefunction implied by the RPA and by the simpler Tamm-Dancoff approximation
(TDA)~\cite{RingSchuck}.  
In the TDA, we consider all allowed single-excitation, one-particle+one-hole
(1p1h) states,
$|\Psi_\mathrm{TDA}(\vq)\rangle = \sum_{\vk}^\prime 
    r_{\vk}^{\vk+\vq} a^\dagger_{\vk+\vq} a_{\vk} |0\rangle$,
where here and throughout the primed summation enforces that $\vk$ is an
occupied state and $\vk+\vq$ is an unoccupied state in the mean-field reference
$|0\rangle$.  Neglecting electron-hole exchange and solving the configuration
interaction problem leads to the identification of the plasmon wavefunction
\begin{equation}
|\Psi_\mathrm{TDA}^\mathrm{P}(\vq)\rangle = N(q) \sum_\vk^\prime 
    \frac{1}{\omega_{\mathrm{TDA}}(q) 
        - (\varepsilon_{\vk+\vq} - \varepsilon_\vk)}
    a^\dagger_{\vk+\vq} a_{\vk} |0\rangle,
\end{equation}
where $N(q)$ is a normalization factor. The TDA plasmon energy
$\omega_{\mathrm{TDA}}(q)$ is the largest root of the secular equation,
\begin{equation}
1 = v(q) \sum_{\vk}^\prime \frac{1}{\omega_{\mathrm{TDA}}(q)
    - (\varepsilon_{\vk+\vq} - \varepsilon_\vk)},
\end{equation}
where $v(q) = 4\pi/q^2$.
The TDA plasmon wavefunction is a coherent superposition of all allowed
1p1h excitations, with equal positive weights in the $q\rightarrow
0$ limit, $N^\prime \rho_\vq |0\rangle$.  
Though physically transparent, the TDA yields a plasmon energy with
an unphysical divergence as $q\rightarrow 0$. This behavior is fixed in the RPA,
which resums ring diagrams in both the forward and backward time directions.
The backward propagations are consistent with correlation in the ground-state
wavefunction. Following the recent result of Ref.~\onlinecite{Berkelbach2018},
the RPA wavefunctions can be
written as a coherent superposition of single excitations on a correlated
ground-state wavefunction $|\Psi_0\rangle = e^{T_2}|0\rangle$,
\begin{equation}
|\Psi_\mathrm{RPA}(\vq)\rangle 
    = \sum_\vk^\prime r_{\vk}^{\vk+\vq} a_{\vk+\vq}^\dagger a_{\vk}
        e^{T_2} |0\rangle,
\end{equation}
where $T_2=\sum_{\vk_1\vk_2\vq}^\prime t_{\vk_1,\vk_2}^{\vq} 
a^\dagger_{\vk_1+\vq} a^\dagger_{\vk_2-\vq}
a_{\vk_2} a_{\vk_1}$
is a double-excitation operator 
with amplitudes $t_{\vk_1,\vk_2}^{\vq}$ satisfying the ring-CCD (RPA)
equations~\cite{Freeman1977,Scuseria2008,Berkelbach2018}.
It is simple to show that the RPA plasmon amplitudes $r_\vk^{\vk+\vq}$
have the same form as those of the TDA,
\begin{equation}
|\Psi_\mathrm{RPA}^\mathrm{P}(\vq)\rangle = N(q) \sum_\vk^\prime 
    \frac{1}{\omega_{\mathrm{RPA}}(q) 
        - (\varepsilon_{\vk+\vq} - \varepsilon_\vk)}
    a^\dagger_{\vk+\vq} a_{\vk} e^{T_2} |0\rangle
\end{equation}
but with the improved RPA plasmon dispersion $\omega_{\mathrm{RPA}}(q)$.
Therefore, we conclude that the RPA plasmon wavefunction is characterized by a
constructive superposition of single excitations on a CCD ground state.

Electronic states with \textit{dominant single-excitation character} are known
to be improved by the inclusion of double excitations corresponding to
two-particle+two-hole (2p2h) configurations.
In periodic EOM-CCSD, the excited-state wavefunctions are given by
\begin{equation}
\begin{split}
&|\Psi_\mathrm{CC}(\vq)\rangle = \Bigg[\sum_\vk^\prime r_{\vk}^{\vk+\vq} 
        a_{\vk+\vq}^\dagger a_\vk \\
    &\hspace{1em} + \sum_{\vk_1,\vk_2,\vk_3}^\prime 
    r_{\vk_1,\vk_2}^{\vk_3,\vk_1+\vk_2-\vk_3+\vq} 
    a^\dagger_{\vk_3} a^\dagger_{\vk_1+\vk_2-\vk_3+\vq}
    a_{\vk_2} a_{\vk_1} \Bigg]\ e^{T_2} |0\rangle,
\end{split}
\end{equation}
and thus include both single and double excitations with respect to the
correlated ground-state wavefunction ($T_2$ is obtained from the full CCSD
equations and $T_1$ is zero for the UEG). Therefore, EOM-CCSD is expected to
provide an improved description of the plasmon, whose entire theoretical
description to date has relied upon a single-excitation picture.
This formalism is equivalent to the diagonalization of the similarity-transformed
Hamiltonian matrix in the basis of 1p1h and 2p2h configurations, as shown
in Fig.~\ref{fig:scheme}.

Detailed expressions for the RPA, TDA, and EOM-CCSD polarizability are given in the
Supplemental Material [].

\textit{Results.}
In Fig.~\ref{fig:sqw_2d_rs4}, we show the dynamic structure factor at $r_s=4$
calculated using the noninteracting theory in the thermodynamic limit (a) as
well as the RPA (b) and EOM-CCSD (c) results for $N=66$ electrons with $M=81$
plane-wave basis functions.  Ignoring symmetries, the EOM-CCSD Hamiltonian
includes more than $2\times10^6$ many-body states.  
Due to the finite system size, in all calculations
we use a broadening of $\eta = 1$~eV.  The simulation data are 
unavailable at large $q$ because of the finite basis set and at small $q$
because of the finite system size.
In particular, for an $N$-electron
simulation, the smallest accessible value of the momentum transfer is $q =
2\pi/L \approx 2.03 q_\mathrm{F}/N^{1/3}$, where $q_\mathrm{F}$ is the Fermi
wavevector; for $N=66$ electrons, this corresponds to $q \approx
q_\mathrm{F}/2$. 
In the Supplemental Material [], we present a thorough study of finite basis
effects and finite size effects at the RPA level, for which large calculations
are affordable; we confirm that the properties studied here are not strongly
affected.

\begin{figure}[t]
\centering
\includegraphics[scale=1.0]{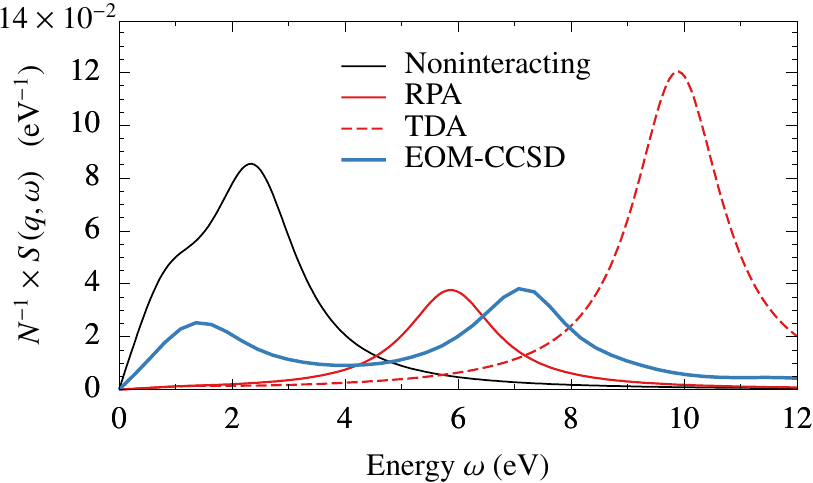}
\caption{
Dynamic structure factor of the UEG at $r_s=4$ with $N=66$ electrons in $M=81$
plane-wave basis functions at $q\approx q_\mathrm{F}/2$. A broadening of $\eta =
1$~eV is used in all calculations.
}
\label{fig:sqw_q0_rs4}
\end{figure}

Despite the finite system size, the plasmon peak at $q=q_\mathrm{F}/2$ is 
separated from the particle-hole continuum and can be confidently assigned
and analyzed.  In Fig.~\ref{fig:sqw_q0_rs4}, we show the dynamic structure
factor at $q=q_\mathrm{F}/2$ calculated using various theories for the
66-electron system.  
The correlated theories are consistent with expected behavior, showing a
redistribution of oscillator strength from the particle-hole continuum into the
plasmon resonance. Interestingly, the EOM-CCSD spectrum maintains a non-negligible
intensity in the low-energy region, which can also be seen in Fig.~\ref{fig:sqw_2d_rs4}(c).  
The energies of the RPA and TDA plasmons are close to their
values in the thermodynamic limit. However, contrary to expectations, the
EOM-CCSD plasmon is located at a \textit{higher} energy than the RPA plasmon, in
disagreement with the known exact behavior.

To understand this behavior, we analyze the wavefunction character of
the states contributing to the plasmon peak.  Unlike the TDA or RPA, for which
the plasmon peak comes from a \textit{single} quantum state, EOM-CCSD predicts a
plasmon peak that is composed of many states, i.e.~it has a physical linewidth
due to interactions with multipair configurations.  This correlation-induced
linewidth is masked by the use of the relatively
large line broadening factor $\eta = 1$~eV.  To characterize
these states, we use an energy-targeting algorithm that locates interior
eigenvalues of the EOM-CCSD Hamiltonian.  Remarkably, we find that 
\textit{all} states contributing to the plasmon peak have a significant
double-excitation (two-partice, two-hole) character; the state
that contributes most strongly to the plasmon peak has only 17\%
single-excitation character and 83\% double-excitation character. This result is
in apparent contrast to the usual picture of the plasmon as a single quantum
state that is well-described as being dominated by single excitations, but
is consistent with a many-body lifetime of such a single quantum state.  The
wavefunction character of the contributing many-body states also explains the
incorrect behavior of the plasmon dispersion: EOM-CCSD is known to overestimate the
excitation energy of states with significant double-excitation character.  A
quantitative prediction of the plasmon energy would require the use of triple
excitations (three-particle, three-hole configurations) to allow
orbital relaxation in the presence of double excitations.

\begin{figure}[b]
\centering
\includegraphics[scale=1.0]{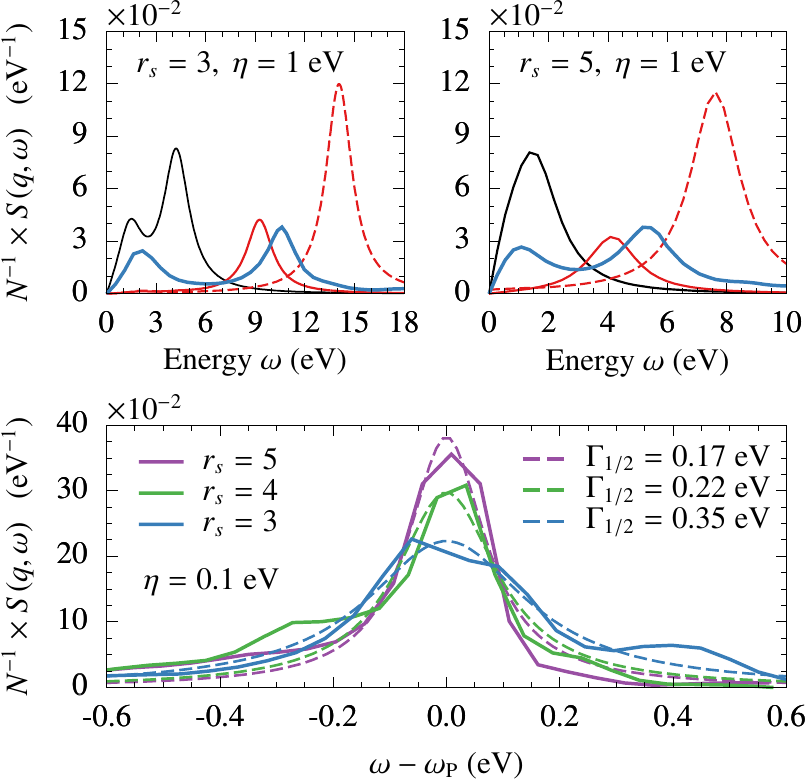}
\caption{
The same as in Fig.~\ref{fig:sqw_q0_rs4}, but for $r_s=3$ (top left)
and $r_s=5$ (top right).
At higher densities (smaller $r_s$), the plasmon peak at 
$q=q_\mathrm{F}/2$ is much closer to the particle-hole continuum,
leading to the enhanced broadening of the EOM-CCSD plasmon, which
can be resolved with a smaller numerical broadening of $\eta=0.1$~eV
(bottom). Dashed lines demonstrate a Lorentzian fit
where $\Gamma_\mathrm{1/2}$ is the half-width at the half-maximum.
}
\label{fig:sqw_q0_lifetime}
\end{figure}

To test these conclusions, we calculated the dynamic structure factor at higher
and lower densities.
In
Fig.~\ref{fig:sqw_q0_lifetime}, we show the dynamic structure factors calculated
at $q=q_\mathrm{F}/2$, for $r_s=3$ and $r_s=5$.  Consistent with our findings at
$r_s=4$, we see that the EOM-CCSD plasmon is situated in between those of the
RPA and TDA, but with a broad lineshape and background due to strong mixing with multipair
excitations.
At higher density (smaller $r_s$), the
importance of the Coulomb interaction is reduced, the RPA is more accurate, and
the wavefunctions are less strongly correlated. Also, the plasmon is shifted to
higher energies. However, the plasmon at the minimum value of $q$
accessible in our 66-electron simulation becomes closer to the particle-hole
continuum and lifetime effects are expected to increase.  At a given
value of $q/q_\mathrm{F}$, the linewidth is roughly proportional to the 
plasmon energy~\cite{Hasegawa1969}, and thus we expect to see an increased plasmon
linewidth for decreasing $r_s$, despite the decrease in electron correlation.
As mentioned above, the physical linewidth is masked by the use of a relatively
large broadening $\eta=1$~eV, which was chosen to minimize finite-size effects in the
particle-hole continuum.  In order to estimate the interaction-induced linewidth,
we re-calculated the dynamic structure factor near the plasmon peak using a much
smaller broadening $\eta=0.1$~eV, shown in the bottom panel of Fig.~\ref{fig:sqw_q0_lifetime}.
With decreasing $r_s$, the plasmon peak acquires significant spectral structure
indicative of contributions from multiple many-electron states.  By fitting to
a Lorentzian lineshape, we extract an approximate plasmon linewidth, which is found
to be $\Gamma_{1/2} =0.17,\ 0.22,\ 0.35$~eV for $r_s=5,\ 4,\ 3$.
We emphasize that the RPA linewidth is precisely the numerical broadening 
$\Gamma_{1/2} = \eta = 0.1$~eV for all values for $r_s$ at this momentum.

Previous diagrammatic calculations on the electron gas have made predictions of
the plasmon linewidth~\cite{Dubois1959,Ninham1966,Dubois1969,Hasegawa1969}.  At leading
order in $q$, the plasmon linewidth at our studied value of $q = q_\mathrm{F}/2$
is given by $\Gamma_{1/2} = b \omega_\mathrm{P}/4$, where $b$ is calculated by
the theory.  The results of Ref.~\onlinecite{Hasegawa1969} are based on a pair
factorization of the four-particle Green's function, $G_4 \approx G_2 \cdot
G_2$. When the noninteracting $G_2$ is used in the factorization, $b$ is on the
order of unity; when the RPA-screened $G_2$ is used in the factorization, $b$ is
significantly reduced to 0.1 or less~\cite{Hasegawa1969}.  Using $b=0.1$ predicts
a linewidth of $\Gamma_{1/2} = 0.10,\ 0.15,\ 0.23$~eV for $r_s=5,\ 4,\ 3$,
in surprisingly good quantitative agreement with our EOM-CCSD results.
We note that the polarizability diagrams responsible for lifetime effects in
EOM-CCSD go beyond the factorization approximation and include interactions
between particle-hole pairs~\cite{Berkelbach2018}, although the screening of such interactions does
not include all time-orderings included in the RPA~\cite{Lange2018}.

\textit{Conclusions.}
We have demonstrated that EOM-CCSD is a promising method for the study of
electronic spectra in condensed-phase systems.  Most significantly, the
inclusion of double excitations in EOM-CCSD enables the ab initio calculation of
interaction-induced lifetimes of quasiparticle resonances.  Here, the
wavefunction-based nature of EOM-CCSD allowed a precise characterization of the
many-body quantum states contributing to the plasmon resonance, which were found
to have significant double-excitation character.

The low single-excitation character of the states contributing to the plasmon
resonance is consistent with the characterization of the plasmon as a 
quasiparticle excitation, i.e.~the quasiparticle weight of the entire plasmon
resonance is conserved, but necessarily shared by the many quantum states
contributing.  This observation has important implications for ab initio
calculations of condensed-phase spectra: the hybridization responsible for
non-negligible electronic linewidths directly implies a large double-excitation
character, which may lead to an overestimation of excitation energies.
This may be responsible for the slight overestimation of the energy of the
plasmon satellite peaks observed in a previous application of EOM-CCSD to
the one-particle spectral function of the UEG~\cite{McClain2016}.
Future work in this direction should pursue the use of triple excitations in
order to realize an ab initio method capable of predicting accurate energies
and lifetimes of condensed-phase quasiparticle excitations.
Comparison to other post-RPA methods, such as the real-time Kadanoff-Baym
approach~\cite{Kwong2000} or the Bethe-Salpeter equation~\cite{Maggio2016} would
also be interesting.
Additionally, studies at larger system sizes and in larger basis sets
will allow the investigation of the modified plasmon dispersion inside
the particle-hole continuum as well as the asymmetry and fine structure
of the dynamic structure factor~\cite{Ichimaru1982}.

The multi-pair nature of the plasmon uncovered here also has a number of
experimental implications,  despite the fact that phonons and interband
scattering can obscure the plasmon's correlation-induced 
lifetime~\cite{Dubois1969,Ku1999}.  In one direction, the physics described here
is potentially important for cold-atom experiments, which provide isolation from
a thermalizing environment and access to electronic relaxation
processes~\cite{Gring2012}.  Additionally, we expect that the physics described
here can be observed in nanomaterials, such as graphene
plasmonics~\cite{Yan2013} or zero-dimensional quantum dots.  In these latter
examples, a phonon bottleneck may inhibit phonon emission~\cite{Urayama2001}, leading plasmons and
excitons to decay into multi-pair excitations, i.e.~an inverse Auger effect.
This phenomenon is at the heart of multiple exciton generation~\cite{Nozik2008}
and singlet exciton fission~\cite{Miyata2019}.

\mbox{}

All calculations were performed with the PySCF software package~\cite{Sun2018},
using resources provided by the University of Chicago Research Computing Center.
This work was supported by the Air Force Office of Scientific Research under
award number FA9550-18-1-0058.
T.C.B.~is an Alfred P.~Sloan Research Fellow.
The Flatiron Institute is a division of the Simons Foundation.

\end{document}